\documentclass[a4paper,11pt]{article}
\usepackage{pos}
\usepackage{subcaption}

\title{Recent results in $b \rightarrow s\,\ell\ell$ transitions at LHCb}

\author*[a]{Juan José Juan Castella}

\onbehalf{on behalf of the LHCb collaboration}

\affiliation[a]{University of Cambridge,\\
  Cavendish Laboratory, Cambridge, United Kingdom}

\emailAdd{jjj33@cam.ac.uk}

\abstract{Rare decays of $B$ hadrons, such as $b\rightarrow s\ell \ell$ processes, provide great avenues to search for the influence of new physics (NP). They are highly suppressed in the Standard Model (SM) and therefore sensitive to NP contributions. In recent years, experimental measurements have shown tension with SM predictions. It is still up for debate whether these stem from genuine NP or an incomplete understanding of the theory. These proceedings focus on the recent efforts of the LHCb collaboration to advance this field. Some of the results covered show that tensions in these processes, like in $B^0 \rightarrow K^{*0}\, \mu^+ \mu^-$, still persist upon higher precision. Others aim to shed some light on the theoretical picture through amplitude analyses of decays like $B^0 \rightarrow K^{*0}\, \mu^+ \mu^-$ and $B^+ \rightarrow K^+ \, \mu^+ \mu^-$. We also present two innovative measurements: the branching fraction of $B^0 \rightarrow K^+ \pi^- \mu^+ \mu^-$ at large dihadron mass; and the first hadron collider measurement of the lepton flavour universality ratio of $B^0 \rightarrow K^{*0} \, \ell^+\ell^-$ at high $q^2$ (dilepton mass squared).}

\FullConference{Presented at the 21st Conference on B-Physics at Frontier Machines, BEAUTY 2026.
\\
1-5 June 2026.
\\
Maastricht, Netherlands.
}

\begin{document}
\maketitle

\section{Introduction}
Experimental efforts on pushing the boundaries of the Standard Model (SM) are split into two fronts: direct searches and indirect searches. The direct avenue aims to produce particles on-shell, which makes it possible to measure their properties. This is essential to understand the nature of the new physics (NP) at play. Its main limitation stems from the fact that the mass of new physics particles that can be produced on-shell is limited by the energy of the collider. On the other hand, indirect searches look for NP contributions that induce deviations from SM predictions. Since the new physics particles are not produced on-shell, the mass which an indirect search is sensitive to is not limited by collision energy. For this reason, they are a good complement to direct searches and can guide the design of future machinery.

\paragraph{}
Decays which are highly suppressed in the Standard Model are particularly suitable for indirect searches, since possible new physics effects will not be buried under the SM contribution. An example is electroweak penguin interactions where the beauty quark changes to a strange quark. These decays, also referred to as $b \rightarrow s\,\ell\ell$ processes, are loop-suppressed in the SM. They are not only interesting for their great sensitivity to NP but also because they give access to a wide range of observables, including branching fractions (BF), charge-parity violating (CPV) observables, angular observables and lepton flavour universality (LFU) ratios. 

\paragraph{}
 Persisting tensions between theory and experiment in these measurements have puzzled the community for some time. An example is the $P_5^{\prime}$ angular observable in $B^0 \rightarrow K^{*0}\, \mu^+ \mu^-$ decays, where measurements across ATLAS \cite{ATLAS_P5}, CMS \cite{CMS_P5}, LHCb \cite{Kstar_ang_old} and Belle \cite{Belle_P5} are all compatible with each other and overall in tension with the SM. Similarly, measurements of the branching fractions of $B^+ \rightarrow K^+ \ell^+ \ell^-$ decays across different experiments consistently lie below the Standard Model prediction \cite{anom_BKll_BaBar,anom_BKll_Belle,anom_BKmumu_LHCb,anom_BKmumu_CDF}. There currently exists no clear explanation behind these anomalies. A real possibility is that these are some of the first pieces of evidence of a NP mechanism at play. Alternatively, the tension could be attributed to the Standard Model predictions themselves. The high-energy or short-distance component of the decay is well understood through perturbative weak effective field theory. However, there are still open questions on the modeling of long-distance contributions from local hadronic form factors and non-local resonance loops ($s\bar{s}$, $c\bar{c}$, $D\bar{D}$, etc.).

\paragraph{}
Progress to resolve these open questions is steadily being made by theory and experimental groups. The following proceedings focus on the recent efforts by the LHCb collaboration in the field of $b \rightarrow s\,\ell\ell$ decays, whose detector allows for world-leading measurements of these processes. We cover five different analyses, starting with an updated measurement of the full set of observables in $B^0 \rightarrow K^{*0}\,\mu^+ \mu^-$ \cite{Kstar_ang}, including $P_5^{\prime}$. Then we outline "unbinned" measurements of the $B^0 \rightarrow K^{*0}\, \mu^+ \mu^-$ \cite{Kstar_un} and $B^+ \rightarrow K^+ \, \mu^+ \mu^-$ \cite{Kmumu_un} decays, which aim to directly measure the contributions from non-local effects. Finally, we present two novel measurements: the BF of $B^0 \rightarrow K^+ \pi^- \mu^+ \mu^- $ at large $m_{K\pi}$ \cite{Kpimumu_largemass}, in a region of phase space unmeasured before; and the LFU ratio in $B^0 \rightarrow K^{*0} \ell^+ \ell^-$ \cite{RKstar} at large squared dilepton mass ($q^2$), measured for the first time at a hadron collider. Most of the analyses discussed utilized the full Run 1 and 2 LHCb datasets excluding the year 2015 and it will be highlighted when otherwise.

\section{Branching fraction and angular analysis of $B^0 \rightarrow K^{*0}\, \mu^+ \mu^-$} \label{sec:Kstar_angular}

The topology of the $B^0 \rightarrow K^{*0}\, \mu^+ \mu^-$ decay makes it a very interesting avenue for probing NP. The $K^{*0}$ meson is reconstructed through its decay to $K^+ \pi^-$, where the spin-parity of the $K^+ \pi^-$ system distinguishes two contributions: the $1^-$ or P-wave and the $0^+$ or S-wave. The full system is parametrized by 5 variables: the dilepton mass ($m_{\mu\mu}$), the dilepton angle ($\theta_{\mu\mu}$), the dihadron mass ($m_{K\pi}$), the dihadron angle ($\theta_{K\pi}$) and the angle between the dilepton and dihadron planes ($\phi$). The decay's complex dynamics allow for the construction of optimized observables that cancel most of the theoretical uncertainties, such as the aforementioned $P_5^{\prime}$.

\paragraph{}
Several improvements are made by this new analysis \cite{Kstar_ang} over the previous result \cite{Kstar_ang_old}. For instance, it increases the amount of data from 4.7 to 8.4 fb$^{-1}$, fits a wider range of the dihadron mass (745.9, 1095.9 $\text{MeV}/c^2$) and considers contributions from non-zero lepton mass. Notably, it also provides a full treatment of S and P-wave components, including their interference, allowing for the first measurement of the full set of CP-averaged and CP-asymmetric observables.

\paragraph{}
The fit to the data is performed in bins of $q^2$ and simultaneously in several variables: $m_{K\pi}$, $\cos{(\theta_{\mu\mu})}$, $\cos{(\theta_{K\pi})}$, $\phi$ and the $B^0$ mass ($m_{K\pi\mu\mu}$). This 5-dimensional fit allows for a simultaneous measurement of the branching fraction and angular observables.

\begin{figure}[htbp] 
  \centering
  \begin{subfigure}[b]{0.48\textwidth}
    \centering
    \includegraphics[width=\textwidth]{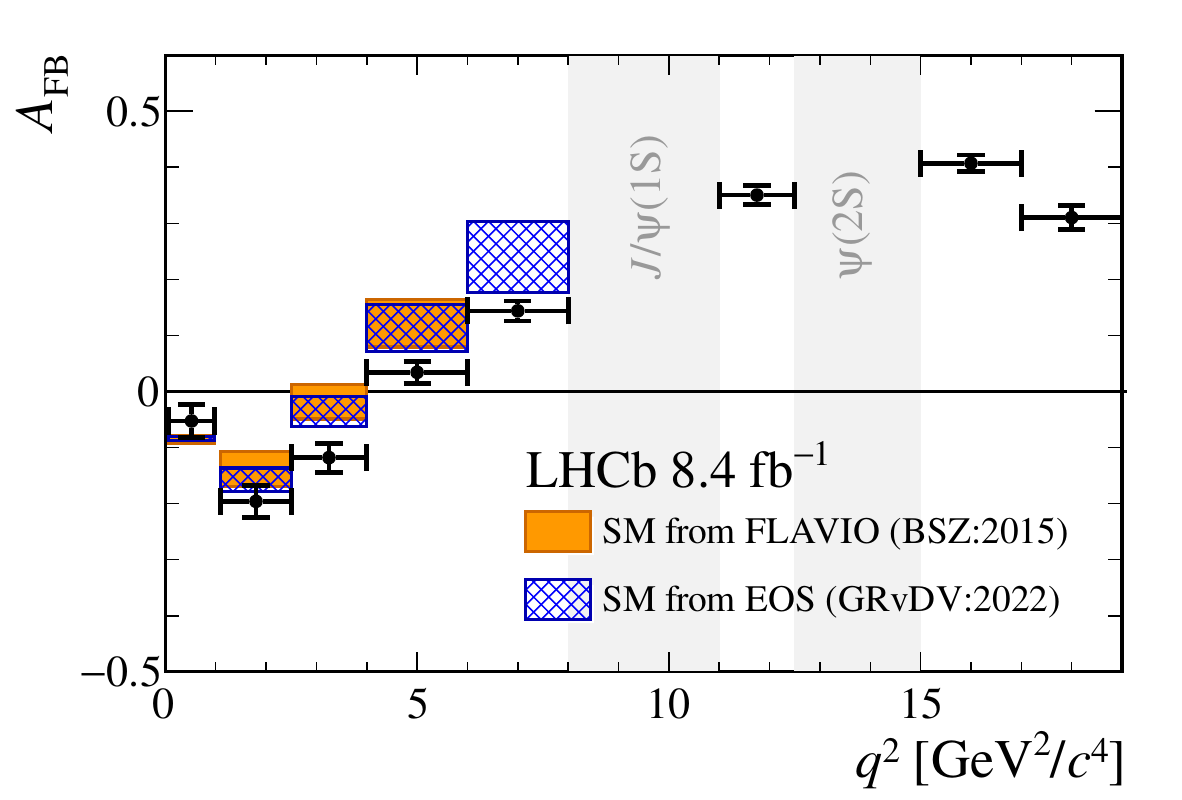}
    \caption{}
  \end{subfigure}
  \hfill
  \begin{subfigure}[b]{0.48\textwidth}
    \centering
    \includegraphics[width=\textwidth]{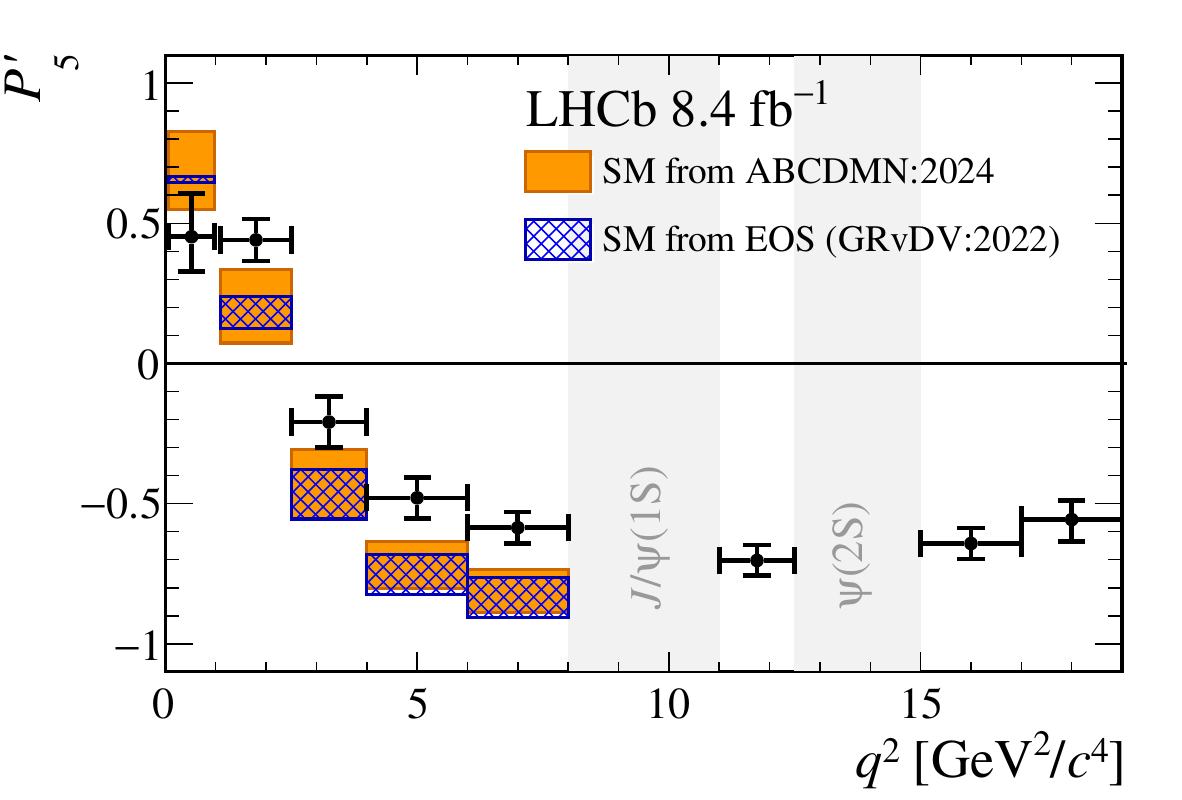}
    \caption{}
  \end{subfigure}

\caption{\label{fig:Kstar_ang} Measurements of the $B^0 \rightarrow K^{*0}\, \mu^+ \mu^-$ forward-backward asymmetry (a) and  $P_5^{\prime}$ (b) in bins of $q^2$ \cite{Kstar_ang}. Standard Model predictions are also shown for the region $q^2 < 8.0 \,\text{GeV}^2/c^4$ \cite{Flavio,BSZ2015,AS2015,EOS,HLMW2015,GKvD2019,Kstar_un_FF1,ABCDMN2023}. The theory predictions seem to be in tension with the measurements particularly near the $J/\psi$ resonance peak.}
\end{figure}

\paragraph{}
Measurements for all CP-violating and most CP-conserving observables are found to agree with the Standard Model. However, as can be seen in figure \ref{fig:Kstar_ang}, $P_5^{\prime}$ and the forward-backward asymmetry ($A_{FB}$) are in tension with the SM predictions \cite{Flavio,BSZ2015,AS2015,EOS,HLMW2015,GKvD2019,Kstar_un_FF1,ABCDMN2023}, particularly in the region of $q^2$ near the $J/\psi$ charmonium resonance. The measurements sit between 1.5 and 2.7 standard deviations away from the SM and are consistent with previous results \cite{Kstar_ang_old}. When fitting the results using the \textsc{Flavio} (\textsc{EOS}) packages \cite{Flavio, EOS}, they produce a 3.6 (3.8)$\,\sigma$ tension with the SM in the real part of the Wilson coefficient $\mathcal{C}_9$. 

\begin{figure}
    \centering
    \includegraphics[width=0.5\linewidth]{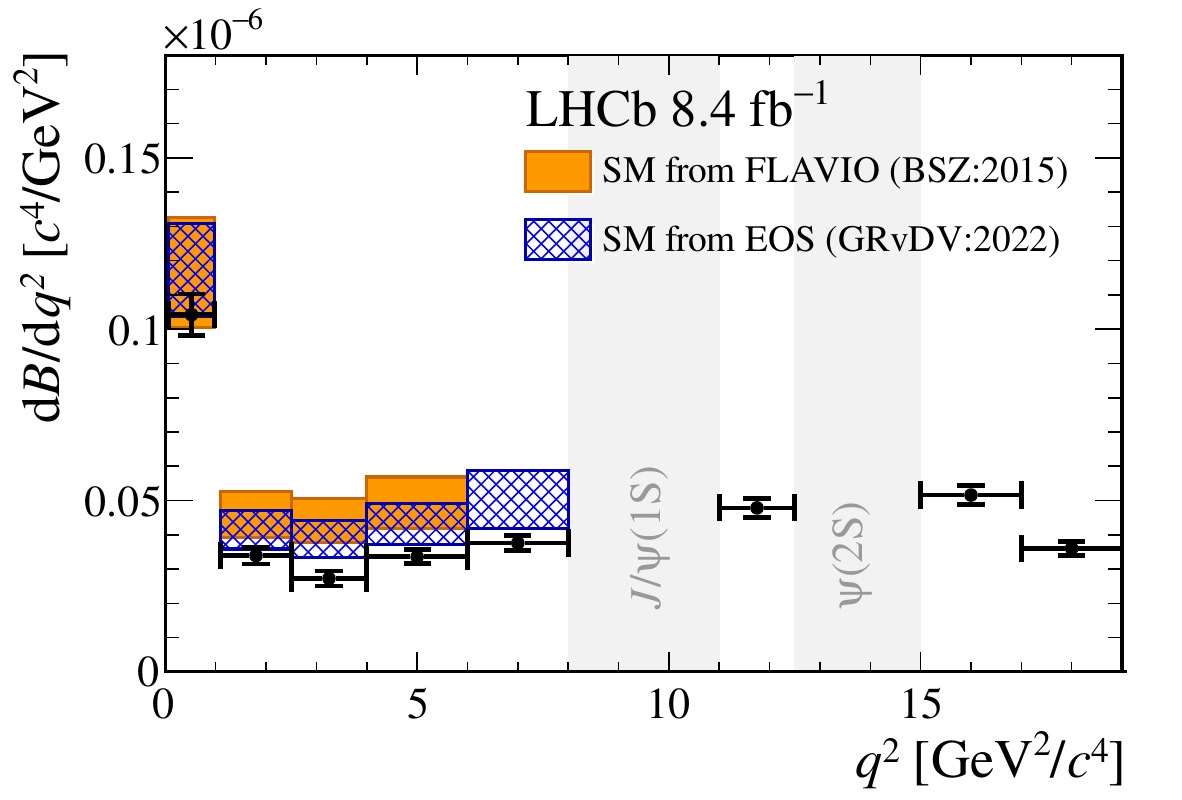}
\caption{\label{fig:Kstar_BR} Measurement of the $B^0 \rightarrow K^{*0}\, \mu^+ \mu^-$ differential branching fraction in $q^2$ \cite{Kstar_ang}. Theory predictions are also shown in the region of $q^2$ below the $J/\psi$ resonance, which seem to overestimate the measurement \cite{Flavio,BSZ2015,AS2015,EOS,HLMW2015,GKvD2019,Kstar_un_FF1}.}
\end{figure}

\paragraph{}
The branching fraction measurement, figure \ref{fig:Kstar_BR}, lies below the SM predictions in the region $1.1 < q^2 < 6.0\,\text{GeV}^2 / c^4$ \cite{Flavio,BSZ2015,AS2015,EOS,HLMW2015,GKvD2019,Kstar_un_FF1}. Including these results in the \textsc{Flavio} (\textsc{EOS}) fits raises the $\Re(\mathcal{C}_9)$ tension to 4.0 (4.1)$\,\sigma$.

\section{Unbinned charge parity violation analysis of $B^0 \rightarrow K^{*0}\, \mu^+ \mu^-$}
Unlike the analysis discussed in section \ref{sec:Kstar_angular} \cite{Kstar_ang}, an unbinned analysis does not measure observables in bins of $q^2$; instead it performs an amplitude fit over the full dilepton mass spectrum. This requires a model for the long-distance contributions (local and non-local) to be assumed. However, it allows the direct extraction of the values of the Wilson coefficients and the amplitude and phase of the non-local long-distance contributions. 

\paragraph{}
This result is an extension of the CP-conserving amplitude analysis of the same decay \cite{Kstar_un_old} and the long-distance contributions are treated in the same manner. Local P-wave form factors are constrained from theory predictions \cite{Kstar_un_FF1} and S-wave form factors are treated as nuisance parameters with a model inspired by Ref. \cite{Kstar_un_FF2}. The non-local contributions are accounted for by replacing $\mathcal{C}_9$ with the effective Wilson coefficient $\mathcal{C}_9^{\text{eff}}$:
\begin{equation}
\mathcal{C}_9^{\mathrm{eff}} = \mathcal{C}_9^{\mathrm{SM}} + \mathcal{C}_9^{\mathrm{NP}} + Y(q^2)
\end{equation}
where $\mathcal{C}_9^{\text{eff}}$ includes contributions from NP ($\mathcal{C}_9^{\text{NP}}$) and $q^2$-dependent non-local contributions ($Y(q^2)$). 

\paragraph{}
The main purpose of this measurement is to extend the framework to include sensitivity to CP-violating new physics. For this reason, complex contributions in the Wilson coefficients $\mathcal{C}_9$ and $\mathcal{C}_{10}$ are allowed in the fit, as opposed to the previous analysis, which assumed them to be real. Both analyses also measure the real parts of the chirality-flipped Wilson coefficients ($\mathcal{C}^{\prime}_9$ and $\mathcal{C}^{\prime}_{10}$), which are zero in the SM.

\begin{figure}
    \centering
    \includegraphics[width=0.5\linewidth]{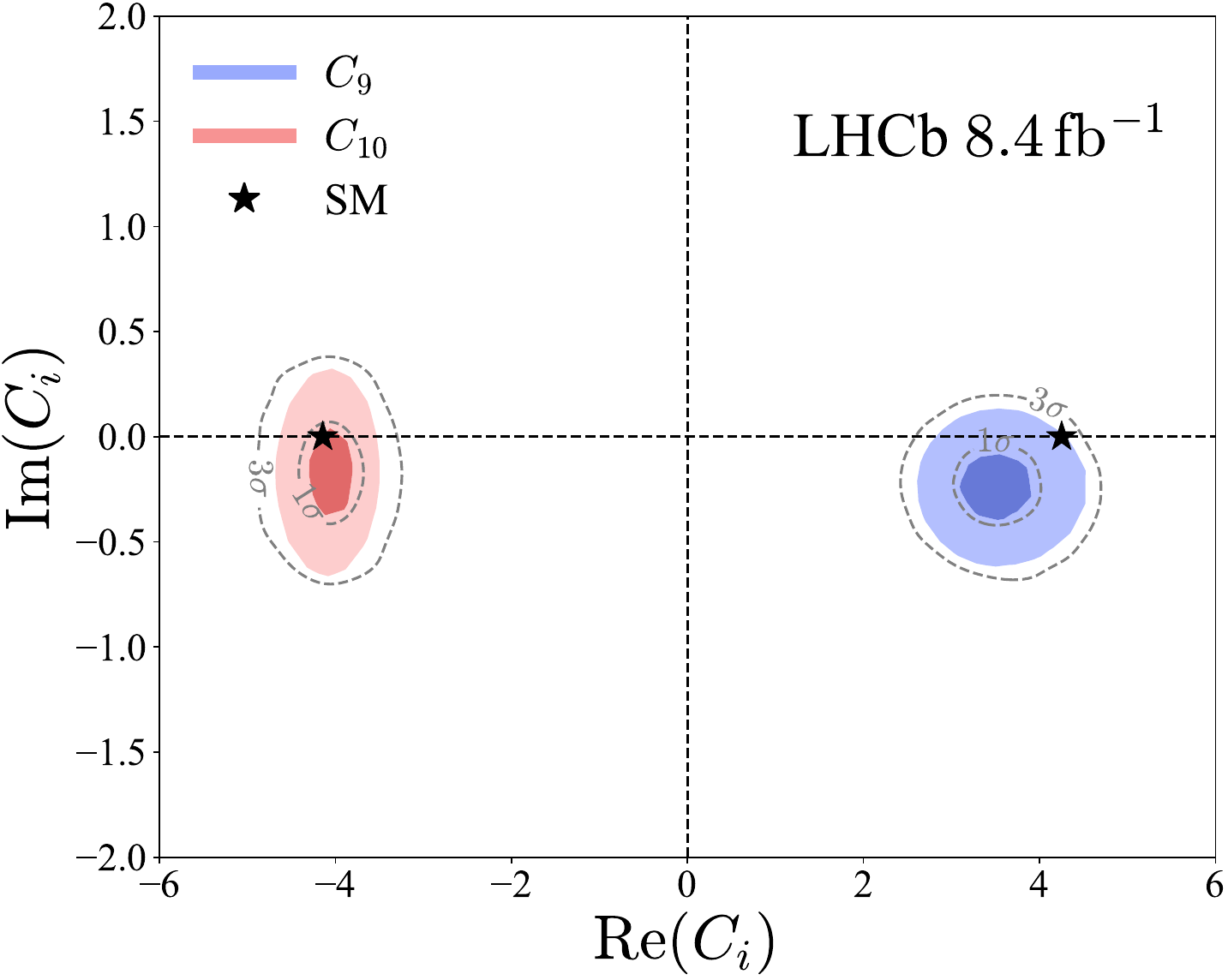}
    \caption{ \label{fig:Kstar_un} Two-dimensional likelihood profiles of $\mathcal{C}_9$ and $\mathcal{C}_{10}$ from the amplitude fit to $B^0 \rightarrow K^{*0}\, \mu^+ \mu^-$ \cite{Kstar_un}. The profile is in the complex plane of the corresponding Wilson coefficient and the other coefficients are not fixed to their SM values.}
\end{figure}

\paragraph{}
The measured $\mathcal{C}_{9,10}$ can be seen in figure \ref{fig:Kstar_un}, where the real and imaginary components are represented as 2D profiles on the complex plane. Note that the other Wilson coefficients ($\mathcal{C}_{10, 9}$, $\Re{(\mathcal{C}^{\prime}_9)}$ and $\Re{(\mathcal{C}^{\prime}_{10})}$) are \textit{not} kept at their SM values. The real parts of the coefficients are compatible with the CP-conserving analysis \cite{Kstar_un_old}, with $\Re(\mathcal{C}_9)$ sitting at 2.1$\,\sigma$ from the SM value. The imaginary parts are measured to a higher precision than the real components; no significant evidence of CP violation is found.

\section{Unbinned analysis of $B^+ \rightarrow K^+ \, \mu^+ \mu^-$}
The amplitude analysis strategy can also be applied to the $B^+ \rightarrow K^+ \mu^+ \mu^-$ decay, potentially shedding some light on the BF anomalies seen in this process. This result, however, is only sensitive to the sum of $\mathcal{C}_9$ and $\mathcal{C}_{10}$ with their chirality-flipped counterparts. These sums are referred to as $\mathcal{C}_V$ and $\mathcal{C}_A$ respectively. 

\begin{figure}
  \begin{subfigure}{0.48\textwidth}
    \includegraphics[width=\textwidth]{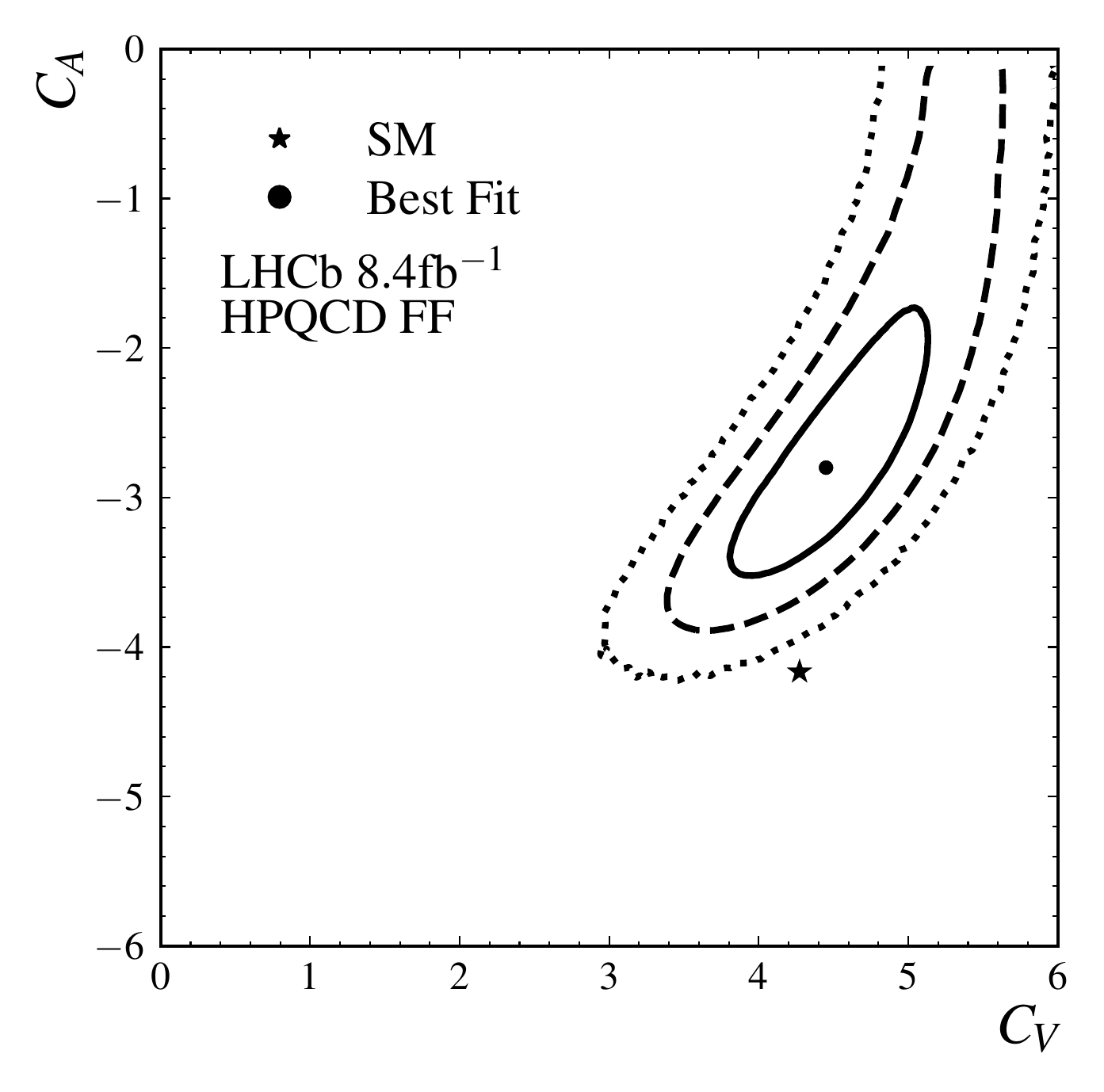}
    \caption{\label{fig:Kmumu_un_2D}}
  \end{subfigure}
  \hfill
  \begin{subfigure}{0.48\textwidth}
    \includegraphics[width=\textwidth]{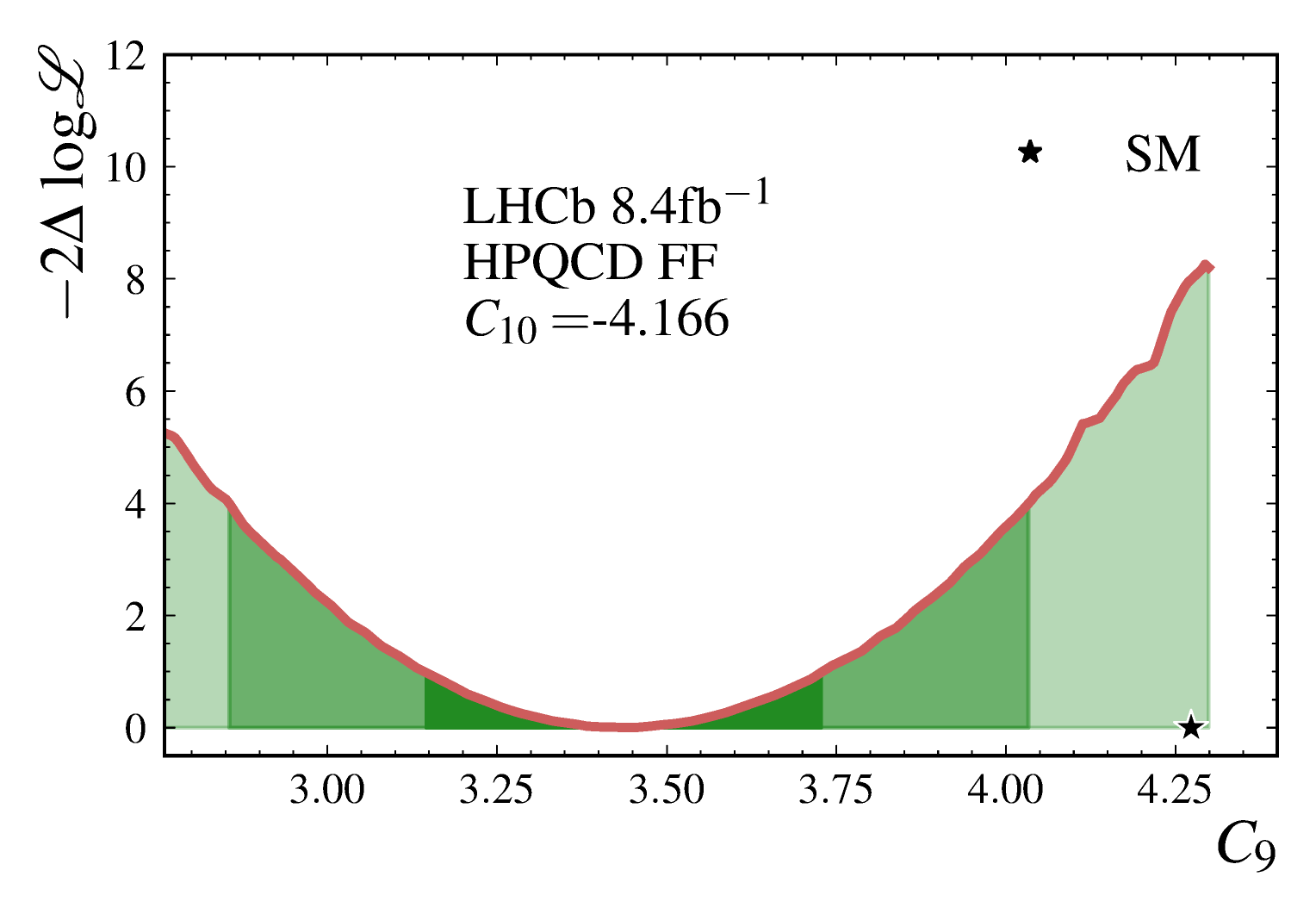}
    \caption{\label{fig:Kmumu_un_1D}}
  \end{subfigure}
\caption{Two-dimensional likelihood profile of the amplitude fit to $B^+ \rightarrow K^+ \mu^+ \mu^-$ using HPQCD form factors (a) \cite{Kmumu_un}. The fit prefers higher than SM values of $\mathcal{C}_A$ and $\mathcal{C}_V$. However, when setting other Wilson coefficients to their SM values, the one-dimensional profile of $\mathcal{C}_{9}$ prefers a value lower than predicted (b). This is compatible with fits to other measurements of $b \rightarrow s \ell \ell $ observables.}
\end{figure}

\paragraph{}
No sensitivity to CPV is included in the fit and therefore the Wilson coefficients are assumed to be real. Two sets of local form factors are tested, those published by the HPQCD \cite{HPQCD} and FNAL/MILC \cite{FNAL} collaborations. Figure \ref{fig:Kmumu_un_2D} shows a 2D likelihood profile of the $\mathcal{C}_A$ and $\mathcal{C}_V$ fit results that used HPQCD form factors. It can be seen that values higher than the SM for both $\mathcal{C}_A$ and $\mathcal{C}_V$ are preferred, which is at odds with other experimental measurements. The overall tension with the Standard Model is $4\,\sigma$ for HPQCD. The FNAL/MILC form factors show a tension in the same direction but of lower strength due to their higher uncertainty ($1.6\,\sigma$). 

\paragraph{}
The one-dimensional likelihood profile of $\mathcal{C}_{9}$ for the HPQCD fit is shown in figure \ref{fig:Kmumu_un_1D}. For this, the coefficients $\mathcal{C}_{10}$, $\mathcal{C}^{\prime}_9$ and $\mathcal{C}^{\prime}_{10}$ are set to their SM values. The $\mathcal{C}_{9}$ tension with respect to the SM now becomes compatible with fits to binned $B^+ \rightarrow K^+ \mu^+ \mu^-$ measurements \cite{Theory_fit_Kmumu} and with other $b\rightarrow s \ell \ell$ results.

\section{Branching fraction of $B^0 \rightarrow K^+ \pi^- \mu^+ \mu^-$ at high dihadron mass}
This measurement is complementary to the  $B^0 \rightarrow K^{*0}(\rightarrow K^+ \pi^-)\,\mu^+ \mu^-$ analysis discussed in section \ref{sec:Kstar_angular} \cite{Kstar_ang}. The branching fraction of the $B^0$ to the $K^+ \pi^- \mu^+ \mu^-$ final state is measured in the region $1000 < m_{K\pi} < 1835\,\text{MeV}/c^2$. This is carried out in bins of $q^2$ over the range $0.1 < q^2 < 18.0 \,\text{GeV}^2/c^4$ and in bins of $m_{K\pi}$ over the range $1.1 < q^2 < 6 \,\text{GeV}^2/c^4$. The only measurement of this BF above the $K^*\,(892)$ resonance pole was performed with LHCb Run 1 data and was limited to a single region $1330 < m_{K\pi} < 1530 \,\text{MeV}/c^2$ and $0.1 < q^2 < 8.0 \,\text{GeV}^2/c^4$ \cite{mkpi_old}.

\begin{figure}
  \begin{subfigure}{0.48\textwidth}
    \includegraphics[width=\textwidth]{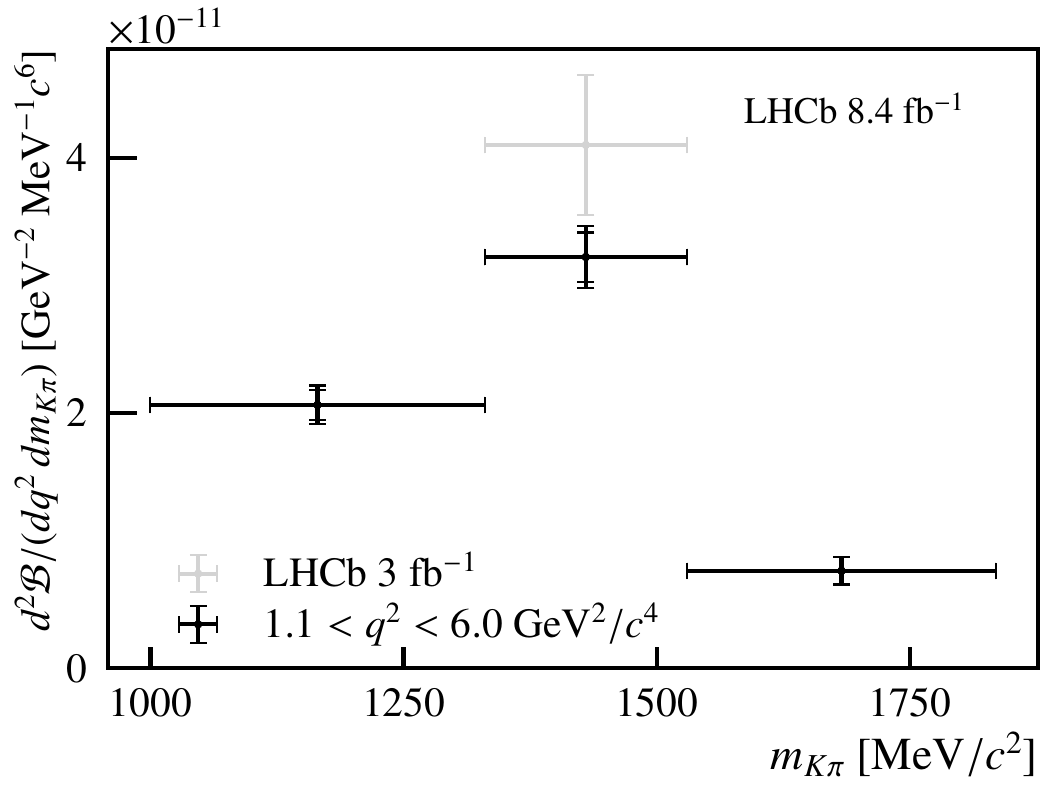}
    \caption{\label{fig:mkpi_a}}
  \end{subfigure}
  \hfill
  \begin{subfigure}{0.48\textwidth}
    \includegraphics[width=\textwidth]{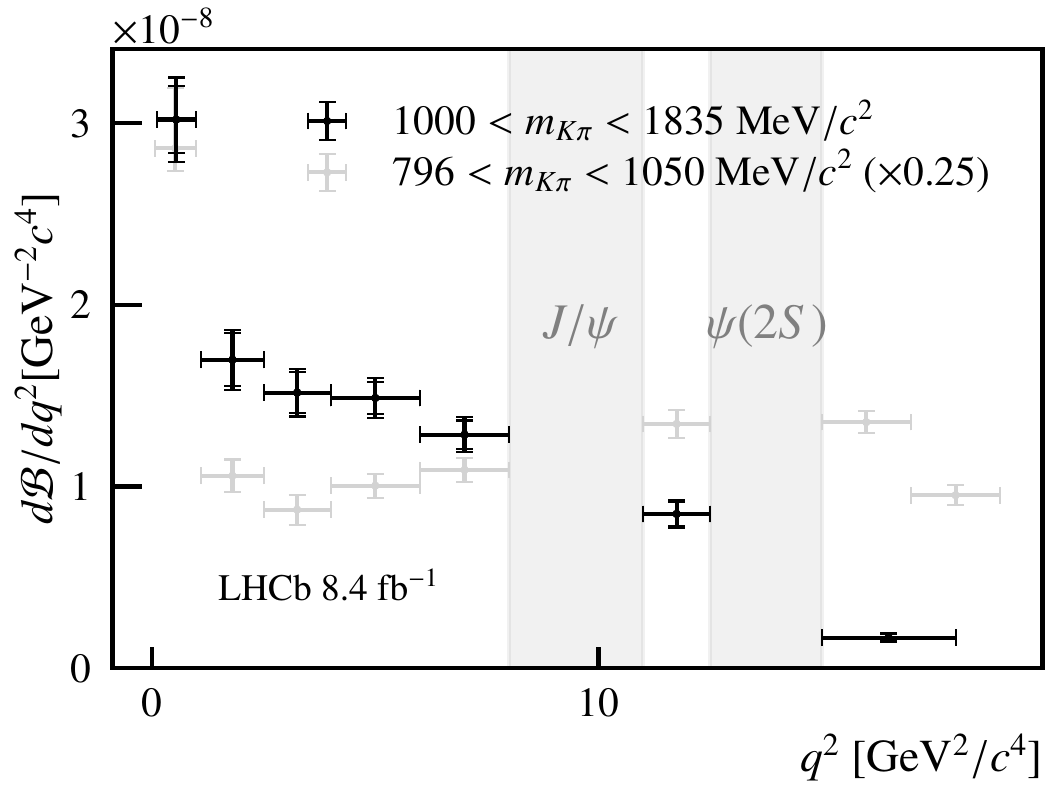}
    \caption{\label{fig:mkpi_b}}
  \end{subfigure}
\caption{Branching fraction of $B^0 \rightarrow K^+ \pi^- \mu^+ \mu^-$ measured at high dihadron mass \cite{Kpimumu_largemass}. The left plot (a) shows the branching fraction in bins of dihadron mass over the region $1.1 < q^2 < 6 \,\text{GeV}^2/c^4$, and the results are compared to the Run 1 LHCb measurement \cite{mkpi_old} in light grey. Figure (b) shows the $q^2$ differential measurement contrasted with the results from \cite{Kstar_ang}, scaled by 0.25 for easier comparison.}
\end{figure}

\paragraph{}
Figure \ref{fig:mkpi_a} compares the dihadron mass differential results to the measurement in \cite{mkpi_old}, integrated over the  $1.1 < q^2 < 6 \,\text{GeV}^2/c^4$ range. The $q^2$ differential measurements are shown in figure \ref{fig:mkpi_b}, together with results from \cite{Kstar_ang} scaled by 0.25. The results of this analysis with those in \cite{Kstar_ang} constitute the first complete determination of the $B^0 \rightarrow K^+ \pi^- \mu^+ \mu^-$  decay width in the $q^2 > 15.0 \,\text{GeV}^2/c^4$ region. This is also the first observation of the $B^0 \rightarrow K^+ \pi^- \mu^+ \mu^-$ above the $K^*\,(892)$ at $q^2$ above the $J/\psi$ meson. There currently are no theory predictions for these measurements.

\section{Lepton flavour universality ratio of $B^0 \rightarrow K^{*0} \ell^+ \ell^-$ at high $q^2$}
Lepton flavour universality ratios serve as great stress tests of the Standard Model, since they allow for a cancellation of most theoretical uncertainties. The ratio $R_{K^*}$, the LFU ratio normalized to the $J/\psi K^*$ mode (equation \eqref{eq:RKstar}), has been measured by the LHCb collaboration before but only in the region of $q^2$ below $6\,\text{GeV}^2/c^4$ \cite{RKstar_lowq2}. Above the $\psi(2S)$ resonance, it has only ever been measured by the Belle collaboration \cite{RKstar_Belle} and this is the first high $q^2$ $R_{K^*}$ result at a hadron collider. 

\begin{equation}\label{eq:RKstar}
R_{K^*} = \frac{N_{\mu\mu,\text{rare}}}{N_{\mu\mu,J/\psi}} \times \frac{N_{ee,J/\psi}}{N_{ee,\text{rare}}} \times \frac{\varepsilon_{\mu\mu,J/\psi}}{\varepsilon_{\mu\mu,\text{rare}}} \times \frac{\varepsilon_{ee,\text{rare}}}{\varepsilon_{ee,J/\psi}}
\end{equation}

\paragraph{}
Where $N$ corresponds to the yield and $\varepsilon$ to the efficiency. The measurement was made in the region $14 < q^2 < 22 \,\text{GeV}^2/c^4$ using the full Run 1 and 2 dataset of LHCb (9 fb$^{-1}$, including 2015). Yields for the electron and muon mode are found to be $125^{+14}_{-16}$ and $1406 \pm 38$ respectively. Their factor of 10 difference stems from the better muon reconstruction efficiency of the LHCb detector. The LFU ratio is measured to be:

\begin{equation}
R_{K^*} = 1.08_{-0.12}^{+0.14} \text{ (stat) } \pm 0.07 \text{ (syst)}
\label{eq:RKstar-measurement}
\end{equation}

The result is fully consistent with the SM and constitutes the world-leading measurement in terms of precision, a threefold improvement over \cite{RKstar_Belle}.


\section{Conclusions and outlook}
The field of rare decays and $b \rightarrow s\,\ell\ell$ is rich with potential for the discovery of new physics. Tensions with the Standard Model have been found by different experiments and corroborated by recent results. Furthermore, experimental efforts are being made to better understand long-distance contributions in these decays, which hopefully will help clarify whether these tensions are a product of our incomplete grasp of the theoretical picture or genuine NP. The LHCb experiment is able to provide competitive measurements that are world-leading in terms of precision and some  of them the first of their kind. 

\paragraph{}
Due to the nature of indirect searches, higher precision is key in providing sensitivity to new physics. Note that all the results highlighted here only exploited the Run 1 and 2 datasets of LHCb. Over the course of Run 3, the LHCb collaboration managed to record roughly three times as much data as recorded in Runs 1 and 2 combined. This means that future measurements will provide significantly greater precision and discovery potential.

\bibliographystyle{refs}
\bibliography{refs}

\end{document}